\documentclass[aps,pra,reprint,groupedaddress,showkeys]{revtex4-2}
\usepackage{amsmath,graphicx,hyperref,verbatimbox,array,float}
\usepackage{indentfirst}
\usepackage{braket}
\usepackage{diagbox}
\usepackage{amssymb}
\usepackage{bbold}
\usepackage{colortbl}
\usepackage{multirow}
\usepackage{subcaption}
\newcommand{\figref}[1]{Fig.~\ref{#1}}

\renewcommand{\eqref}[1]{Eq.~\ref{#1}}
\newcommand{\tr}{{\mathrm{Tr}}}
\newcommand{\centered}[1]{\begin{tabular}{l} #1 \end{tabular}}

\begin{document}

\title{Quantum tomography of three-qubit generalized Werner states}
\author{Artur Czerwinski}
\email{aczerwin@umk.pl}
\affiliation{Institute of Physics, Faculty of Physics, Astronomy and Informatics \\ Nicolaus Copernicus University,
Grudziadzka 5, 87--100 Torun, Poland}

\begin{abstract}
In this article, we introduce a numerical framework for quantum tomography and entanglement quantification of three-qubit generalized Werner states. The scheme involves the single-qubit SIC-POVM, which is then generalized to perform three-qubit measurements. We introduce random errors into the scheme by imposing the Poisson noise on the measured photon counts. The precision of state estimation is quantified and presented on graphs. As a special case, we compare the efficiency of the framework for the pure three-qubit generalized Werner state and the W state. The reconstructed states are presented graphically and discussed.
\end{abstract}

\maketitle

\section{Introduction}

Quantum state tomography (QST) plays a crucial role in many quantum information protocols since it provides a variety of techniques to determine a mathematical representation of a quantum system \cite{paris04}. Photonic tomography, which is a specific type of QST, has been developed to characterize the quantum state of photons, including polarization \cite{dariano03}. Since noise and errors are inherent to any measurement, one needs to implement statistical methods which produce reliable estimates of the actual states \cite{Hradil1997,Opatrny1997}. Estimation methods can be compared in terms of their efficiency in state reconstruction \cite{Acharya2019}.

Entangled states, which feature non-classical correlations, are an important resource for quantum communication and computing \cite{Einstein1935,Bell1964}. In particular, entangled photon pairs can be used for multiple applications, such as quantum key distribution (QKD) \cite{Ekert1991}. Other well-known applications of entangled states relate to: superdense coding \cite{Bennett1992,Mattle1996}, quantum teleportation \cite{Bennett1993,Anderson2020}, quantum computing \cite{Jozsa2003,Obrien2007}, quantum interferometric optical lithography \cite{Boto2000}, etc. Entanglement in photonic states is not limited only to photon pairs, but can be extended to three-qubit polarization entanglement \cite{Hamel2014}. Therefore, the ability to characterize multipartite entangled states based on measurements remains a relevant issue. This work is a generalization of Ref.~\cite{Czerwinski2021}, where two-qubit Werner states were considered.

In this work, we follow the generalized approach to quantum measurement, which relies on positive operator-valued measures (POVMs) \cite{Nielsen2000}. More specifically, in our QST framework, we apply a symmetric, informationally complete, positive operator-valued measure (SIC-POVM) \cite{Renes2004}. Additionally, we impose the Poisson noise to make the measurement scheme realistic. For different scenarios, we numerically test the model on three-qubit generalized Werner states. Separately, we compare the estimation of GHZ and W states. The accuracy of the framework is quantified and presented on graphs. Also, we introduce and discuss a simplified entanglement measure to quantify the amount of entanglement detected by the measurement scheme.

The paper is organized as follows: in Sec.~\ref{methods}, we introduce in detail the framework for QST of three-qubit states. Then, in Sec.~\ref{results}, the results are presented and discussed. The main conclusions are provided in the summary.

\section{Framework for state tomography}\label{methods}

\subsection{Three-qubit generalized Werner states}

There are many celebrated classes of entangled states that are studied within quantum information theory. For instance, R. Werner introduced in 1989 a one-parameter class of mixed quantum states that feature non-classical correlations \cite{Werner1989}. In particular, two-qubit Werner states have been implemented for various important advancements in quantum information \cite{Lee2000,Yeo2002,Barbieri2004,LiHui2015}. In this work, we analyze three-qubit generalized Werner states \cite{Dur2000,Siewert2012}:
\begin{equation}\label{w4}
\rho_W^{3q} (\eta) = \eta \ket{\mathrm{GHZ}}\!\bra{\mathrm{GHZ}} + \frac{1- \eta}{8} \mathbb{I}_8,
\end{equation}
where $\ket{\mathrm{GHZ}} = (\ket{000} + \ket{111})/\sqrt{2}$ and $0 \leq \eta \leq 1$. The definition means that $\rho_W^{3q} (\eta)$ is a mixture of a Greenberger-Horne-Zeilinger (GHZ) state \cite{Greenberger19990,Zukowski1998} and the completely unpolarized state (i.e., maximally mixed state). The GHZ state can be obtained from three polarization-entangled spatially separated photons \cite{Bouwmeester1999}, which makes it suitable for our framework. This kind of entangled state is famous for its importance in quantum information, including quantum teleportation \cite{Karlsson1998}, quantum secret sharing \cite{Hillery1999}, or quantum cryptography \cite{Kempe1999}. Nonetheless, GHZ states remain a topic of intensive research, see, e.g., Ref.~\cite{Sawicki2014,Zhao2021}.

Since for $\eta = 1$ the density matrix \eqref{w4} represents the GHZ state, we pay special attention to this case. In our QST framework, we compare the accuracy of GHZ state reconstruction with another three-qubit entangled state defined as:
\begin{equation}\label{w5}
\ket{W} = \frac{1}{\sqrt{3}} \left ( \ket{001}  + \ket{010}  + \ket{100}   \right),
\end{equation}
which is commonly referred to as the W state. $\ket{GHZ}$ and $\ket{W}$ are representatives of two very different classes of three-particle entangled states \cite{Dur2000a}, which cannot be transformed into each other by local operations. These states are often compared in the context of three-photon entanglement as well as in a reference to their algebraic properties \cite{Sawicki2013}. The W state was proposed as a resource for several applications, such as secure quantum communication \cite{Jian2007}. Polarization-entangled W states can be obtained by parametric down-conversion from a single-photon source \cite{Yamamoto2002}, which implies that it is applicable in the framework considered in the present article.

\subsection{Measurements}

To obtain information required for QST of three-qubit generalized Werner states, we implement the SIC-POVM for the Hilbert space such that $\dim \mathcal{H} =2$, which can be generalized to three-qubit measurement operators. In case of qubits, the SIC-POVM is constructed from four vectors:
\begin{equation}\label{eqm1}
\begin{split}
&\ket{\xi_1} = \ket{0} \hspace{1cm} \ket{\xi_2} = \frac{1}{\sqrt{3}} \ket{0} + \sqrt{\frac{2}{3}} \ket{1} \\
&\ket{\xi_3} = \frac{1}{\sqrt{3}} \ket{0} + \sqrt{\frac{2}{3}} e^{i \frac{2 \pi}{3}} \ket{1}\\
& \ket{\xi_4} = \frac{1}{\sqrt{3}} \ket{0} + \sqrt{\frac{2}{3}}  e^{i \frac{4 \pi}{3}} \ket{1},
\end{split}
\end{equation}
where by $\{\ket{0},\ket{1}\}$ we denote the standard basis in $\mathcal{H}$. From the definition of the SIC-POMC, the measurement operators are represented as rank$-1$ projectors:
\begin{equation}\label{eqm2}
\mathcal{P}_k := \frac{1}{2} \ket{\xi_k}\! \bra{\xi_k},
\end{equation}
which satisfy $\sum_{k=1}^4 \mathcal{P}_k = \mathbb{I}_2$. The set of four operators $\{\mathcal{P}_k\}$ is considered an optimal measurement scheme for qubit tomography \cite{Rehacek2004}. Moreover, the SIC-POVM is a minimal informationally complete set of measurement operators. However, it proved to be a reliable framework for single-qubit tomography. For example, the measurement represented by the SIC-POVM can be effectively realized on photons to characterize the polarization state of light \cite{Bent2015}. 

In this framework, we assume that the three-qubit state is encoded in the polarization mode of photons. Each photon travels in a separate arm of the experimental setup and undergoes measurements according to the SIC-POVM scheme. Therefore, for state reconstruction of three photons, we propose three-qubit operators:
\begin{equation}\label{eqm4}
M^{3q}_{\kappa} := \mathcal{P}_i \otimes \mathcal{P}_j \otimes \mathcal{P}_k,
\end{equation}
where $i,j,k = 1, \dots, 4$ and, for simplicity, we used one symbol to denote the three-qubit operators, i.e. $\kappa \equiv (i,j,k)$. Naturally, from the definition \eqref{eqm4}, we have $64$ three-qubit measurement operators. This measurement scheme is minimal as far as three-qubit states are concerned. Therefore, it appears intriguing to investigate the performance of such measurements in a noisy scenario.

\subsection{Methods of state reconstruction}

The research goal is to analyze the performance of the measurement scheme based on the SIC-POVM in the process of quantum state estimation. In reality, experimental errors are unavoidable. Therefore, in this model, we consider the Poisson noise (shot noise), which is commonly taken into account as a source of uncertainty in frameworks that involve single-photon counting, see, e.g. Ref.~\cite{Hernandez2007,Hasinoff2014,Shin2015}.

Photon counting is a classic Poisson process since individual photon detections can be treated as independent events that follow a random distribution. Therefore, the photon counts can be described by a standard Poisson distribution characterized by one parameter that expresses both the mean and the variance. Then, by $\mathcal{N}$ we denote the average ensemble size, i.e. the mean number of quantum systems produced by the source per measurement provided the time interval of the detection is fixed. In the three-qubit framework, we assume that the detectors receive coincidence counts, $n^{3q}_{\kappa}$, which can be represented as:
\begin{equation}\label{eqm5}
n^{3q}_{\alpha} = \widetilde{\mathcal{N}}_{\kappa} \,\tr \left(M^{3q}_{\kappa} \, \rho_W^{3q} (\eta)  \right),
\end{equation}
where $\widetilde{\mathcal{N}}_{\kappa} \in \mathrm{Pois} (\mathcal{N})$, i.e., the ensemble size for each act of measurement is selected randomly from the Poisson distribution characterized by the mean value $\mathcal{N}$, cf. Ref.~\cite{Thew2002}. In experimental realizations of QST framework, the value of $\mathcal{N}$ is never known precisely. However, one should bare in mind that the number of photons per measurements varies due to random fluctuations. Mathematical modeling of experimental results \eqref{eqm5} allows us to numerically generate noisy data for any three-qubit Werner state $\rho_W^{3q} (\eta)$.

We assume that the quantum state that has to be identified remains completely unknown prior to measurements. Consequently, the expected photon counts can be modeled by
\begin{equation}\label{eqm6}
m^{3q}_{\kappa} = \mathcal{N} \,\tr \left(M^{3q}_{\kappa} \, \sigma^{3q} \right),
\end{equation}
where $\sigma^{3q}$ denotes a $8 \times 8$ density matrix that can relate to any three-qubit quantum state. The unknown density matrix can be decomposed based on the Cholesky factorization:
\begin{equation}\label{eqm7}
\sigma^{3q} = \frac{T^{\dagger} T}{ \tr (T^{\dagger} T)}
\end{equation}
and $T$ stands for a lower-triangular matrix that involves $64$ real parameters: $t_1, \dots, t_{64}$, cf. \cite{James2001,Altepeter2005}. In multiple QST frameworks, the Cholesky factorization is implemented since it guarantees that the matrix obtained from the scheme is physical. In other words, $\sigma^{3q}$ is Hermitian, positive semi-definite, of trace one.

By following \eqref{eqm7}, we reformulate the problem of state estimation. The unknown quantum state $\sigma^{3q}$ is identified if we determine the parameters $t_1, \dots, t_{64}$ that completely characterize $T$. We implement the $\chi^2-$estimation to optimally match the set of parameters to the noisy data generated numerically, see, e.g., Ref.~\cite{Jack2009}. As a result, we need to determine the minimal value of the following function:
\begin{equation}\label{eqm8}
\chi^2 (t_1, \dots, t_{64}) = \sum_{\kappa=1}^{64} \frac{\left(n^{3q}_{\kappa} -  m^{3q}_{\kappa}\right)^2}{m^{3q}_{\kappa}}.
\end{equation}

The framework consists of specific steps that enable reconstruction of any state $\rho_W^{3q} (\eta)$. Firstly, we numerically generate noisy measurement results according to \eqref{eqm5}. Then, the quantum state is reconstructed by learning the parameters $t_1, \dots, t_{64}$ that correspond to the minimal value of $\chi^2$. We can proceed analogously for other three-qubit states. In particular, we implement the same scheme to the W state given in \eqref{w5}.

\subsection{Performance analysis}

The performance of the framework in the presence of Poisson noise needs to be investigated by introducing appropriate figures of merit. We define three indicators that can be used to analyze and discuss both precision of state estimation and entanglement detection.

Most of all, for each input state $\rho_W^{3q} (\eta)$, we compute its fidelity with the result of $\chi^2-$estimation, $\sigma^{3q}$. By following the standard definition of the quantum fidelity we obtain \cite{Uhlmann1986,Jozsa1994,Bengtsson2006}:
\begin{equation}\label{eqm9}
\mathcal{F}  (\sigma^{3q}, \rho_W^{3q}  (\eta)) := \left(\tr \sqrt{\sqrt{\sigma^{3q}} \, \rho_W^{3q} (\eta)  \, \sqrt{\sigma^{3q}}} \right)^2.
\end{equation}

This figure of merit is commonly utilized to evaluate the precision of QST frameworks. Particularly, the fidelity is useful to discuss the quality of state estimation under conditions that involve experimental noise, see, e.g., Ref.~\cite{Rosset2012,Yuan2016,Titchener2018,Czerwinski2021}. In our model, the fidelity is treated as a function of $\eta$ and, for simplicity, it is denoted by $\mathcal{F}(\eta)$. With this definition, we can track the accuracy of the framework for the class of the three-qubit generalized Werner states.

Secondly, we can compute the purity of the states obtained from the framework. This indicator allows one the evaluate the mixedness of the estimation results. We implement the standard notion of purity, which is denoted by $\gamma$ \cite{Nielsen2000}. More specifically, for each $\sigma^{3q}$, we compute the purity as: $\gamma \equiv \tr (\sigma^{3q})^2$. Again, the purity can be regarded as a function of $\eta$ and denoted by $\gamma (\eta)$. For different values of $\mathcal{N}$, the purity of estimated states can be compared with the actual value corresponding to the input state.

Finally, we implement a simplified measure to quantify the amount of entanglement that can be detected by the measurement scheme. Since there is no universal entanglement measure for three-qubit states, we further exploit the notion of quantum fidelity to approximately compare different states. For $\eta =1$, the three-qubit generalized Werner state is equivalent to the GHZ state, which makes it the most entangled state in this class. Therefore, it appears rational to treat the fidelity between $\sigma^{3q}$ and the GHZ state as an indicator of entanglement:
\begin{equation}\label{eqm9}
\mathcal{F}  (\sigma^{3q},\rho_{GHZ}) := \left(\tr \sqrt{\sqrt{\sigma^{3q}} \,\rho_{GHZ}  \, \sqrt{\sigma^{3q}}} \right)^2,
\end{equation}
where $\rho_{GHZ} = \ket{\mathrm{GHZ}}\!\bra{\mathrm{GHZ}}$. Analogously, we can treat this figure as a function of $\eta$ to track the entanglement preservation along the complete class. Such function will be denoted as $\mathcal{F}_{GHZ} (\eta)$.

\section{Results and analysis}\label{results}

We consider three measurement scenarios that differ in the number of quantum systems provided by the source per measurement (ensemble size). To be more specific, we assume that $\mathcal{N} = 10$, $\mathcal{N} =100$, and $\mathcal{N} =1\,000$. This approach allows us to investigate the performance of the framework versus the amount of Poisson noise which is strictly correlated with the average number of systems involved in one measurement.

\subsection{Tomography of three-qubit generalized Werner states}

\begin{figure}[h]
	\centering
         \centered{\includegraphics[width=0.95\columnwidth]{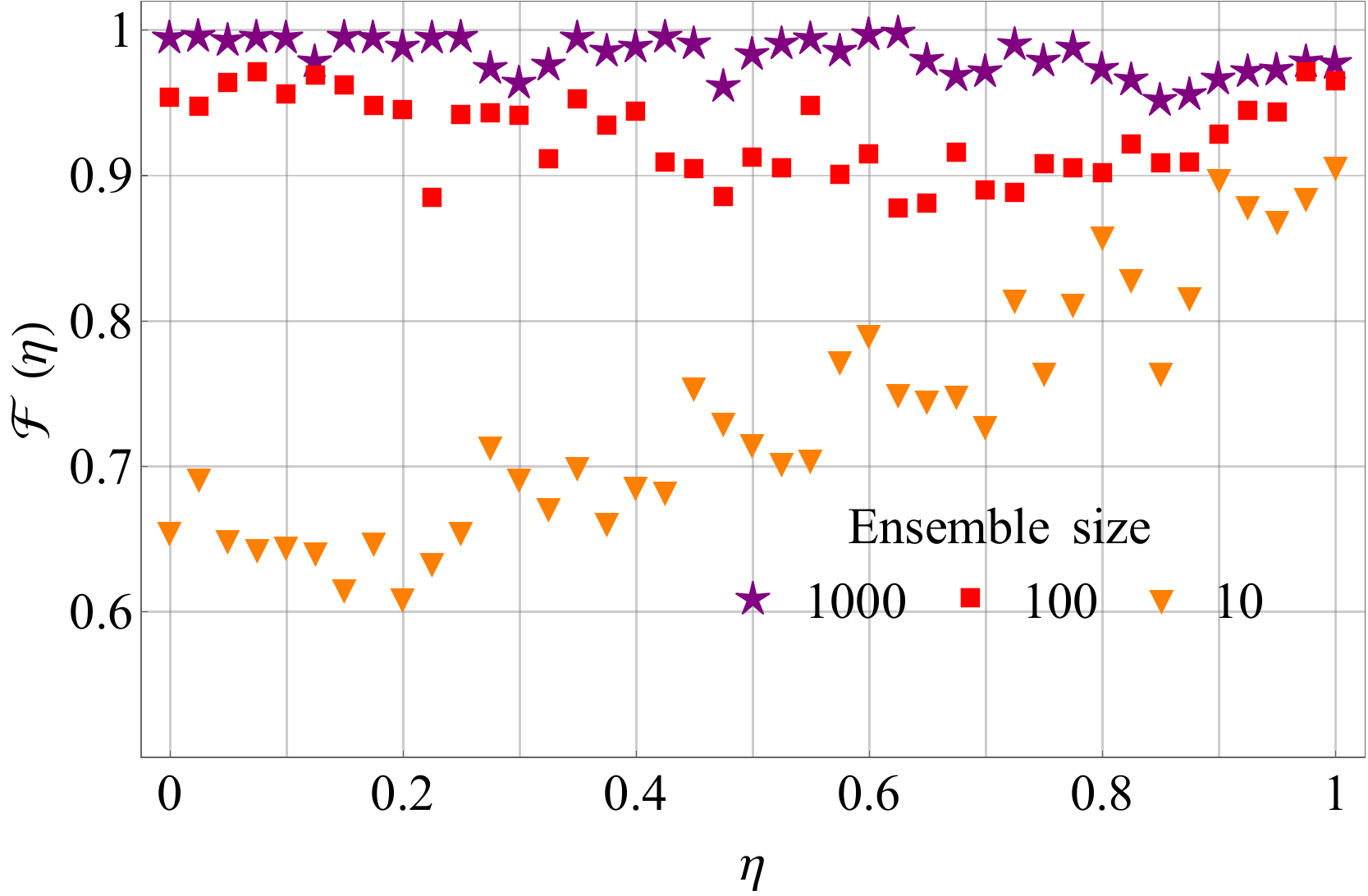}}
	\caption{Plots present the fidelity, $\mathcal{F} (\eta)$, in QST of three-qubit generalized Werner states.}
	\label{fidelity}
\end{figure}

For the three values of $\mathcal{N}$, we perform QST along the entire domain of $\eta$. In \figref{fidelity}, one can observe the plots of the quantum fidelity, $\mathcal{F}(\eta)$. First, for $\mathcal{N} = 1\,000$, one can observe high quality of quantum state estimation. We notice that $\mathcal{F}(\eta) > 0.95$ for all $\eta$. This confirms that the Poisson noise has only a limited impact on state estimation if we utilize a relatively high number of quantum systems per measurement. Although the framework is based on the minimal set of measurement operators, it is robust against the noise in this scenario.

Next, if $\mathcal{N} = 100$, we observe a moderate impact of the Poisson noise on the state tomography. The accuracy of QST remains at a decent level, i.e., $\mathcal{F}(\eta) > 0.87$ for all $\eta$. The plot lies completely below the one corresponding to $\mathcal{N} = 1\,000$. Additionally the distribution of results is irregular, which is caused by the random nature of the noise introduced into the measurement scheme.

Finally, for $\mathcal{N} = 10$, we witness a significant decline in the quality of state recovery. As expected, the Poisson noise has a damaging impact on the quantum fidelity if we utilize only $10$ systems per measurement. We notice that the accuracy of state estimation irregularly goes up and down, reflecting the randomness of noise. Nevertheless, a conclusion can be formulated that $\mathcal{F}(\eta)$ improves as we increase $\eta$. In other words, for the smallest ensemble size, we observe a correlation between the accuracy of state reconstruction and the purity of the input state. This implies that the scenario with $\mathcal{N} = 10$ performs better if the three-qubit generalized Werner state is close to pure.

\begin{figure}[h]
	\centering
         \centered{\includegraphics[width=0.95\columnwidth]{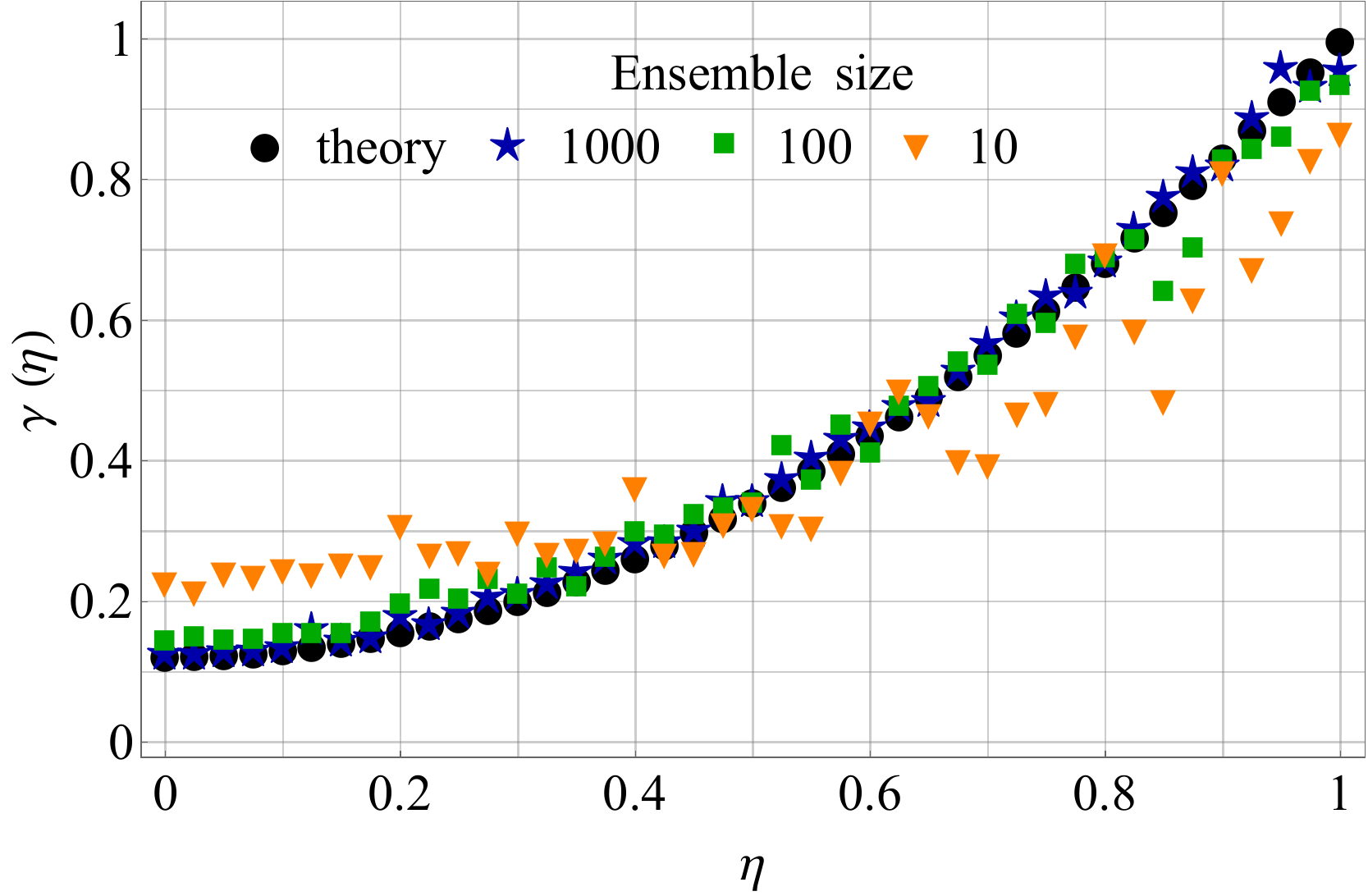}}
	\caption{Plots present the purity, $\gamma (\eta)$, in QST of three-qubit generalized Werner states.}
	\label{purity}
\end{figure}

From a practical point of view, the findings included in \figref{fidelity} imply that we can efficiently implement the framework for the entire class of three-qubit generalized Werner states provided we utilize many copies of the system per measurement. With single systems at our disposal, one can accurately reconstruct only those Werner states that are close to pure.

In \figref{purity}, one finds the purity of the estimated states $\sigma^{3q}$ compared with the actual purity the three-qubit generalized Werner states $\rho_W^{3q} (\eta)$. The plot confirms the previous observations. For $\mathcal{N} = 1\,000$, the plot of $\gamma (\eta)$ strictly overlaps with the theoretical value. If $\mathcal{N} = 100$, we detect minor discrepancies between the purity of the estimates and the actual value. Furthermore, when $\mathcal{N} = 10$, we notice that the plot departs significantly from the theoretical curve. Interestingly, in the case of purity, the estimated figures coincide with the actual values for a middle range of $\eta$, i.e., $ 0.4<\eta <0.8$. For the smallest ensemble size, the plot of the purity first goes above the theoretical line, and then below. This confirms that few quantum systems per measurement can lead to significant inaccuracies.

\begin{figure}[h]
	\centering
         \centered{\includegraphics[width=0.95\columnwidth]{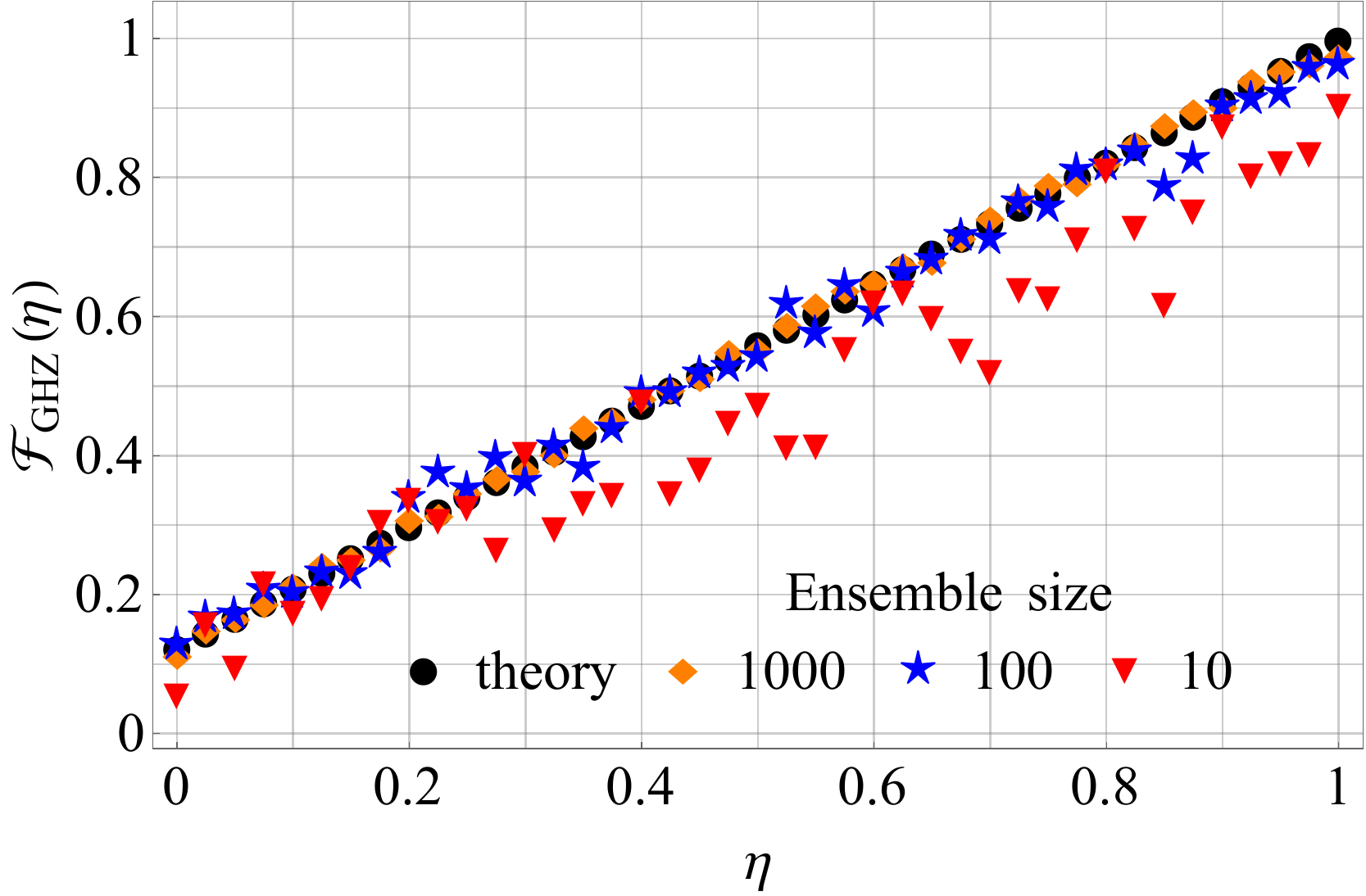}}
	\caption{Plots present the fidelity of reconstructed states with the GHZ state, $\mathcal{F}_{GHZ} (\eta)$, in QST of three-qubit generalized Werner states.}
	\label{entanglement123}
\end{figure}

Finally, we can consider the simplified entanglement measure as defined in \eqref{eqm9}. For actual three-qubit generalized Werner states \eqref{w4}, the function $\mathcal{F}_{GHZ} (\eta)$ grows linearly. We compare this theoretical value with the results corresponding to the estimated states obtained from the framework. In \figref{entanglement123}, one can notice  correlations between the plots for three different values of $\mathcal{N}$. The tendencies are in line with prior figures of merit since for $\mathcal{N} = 1\,000$ or $\mathcal{N} = 100$, we obtain a satisfying overlap between the theoretical path and the results of simulations. For $\mathcal{N} = 10$, we notice significant discrepancies as most points lie below the theoretical path. These results demonstrate that this number of systems per measurement is not sufficient to capture the entire amount of entanglement in three-qubit generalized Werner states.

\subsection{Tomography of GHZ state and W state}

	\begin{figure*}[ht!]
		\centering
		\begin{tabular}{|c|c|c|}
\hline
			\backslashbox[18mm]{\color{black}$\mathcal{N}$}{\color{black}state}&\centered{$\ket{GHZ}$}&\centered{$\ket{W}$}\\\hline
			\centered{$1000$}& \begin{tabular}{l}\raisebox{- \height}{\includegraphics[width=0.65\columnwidth]{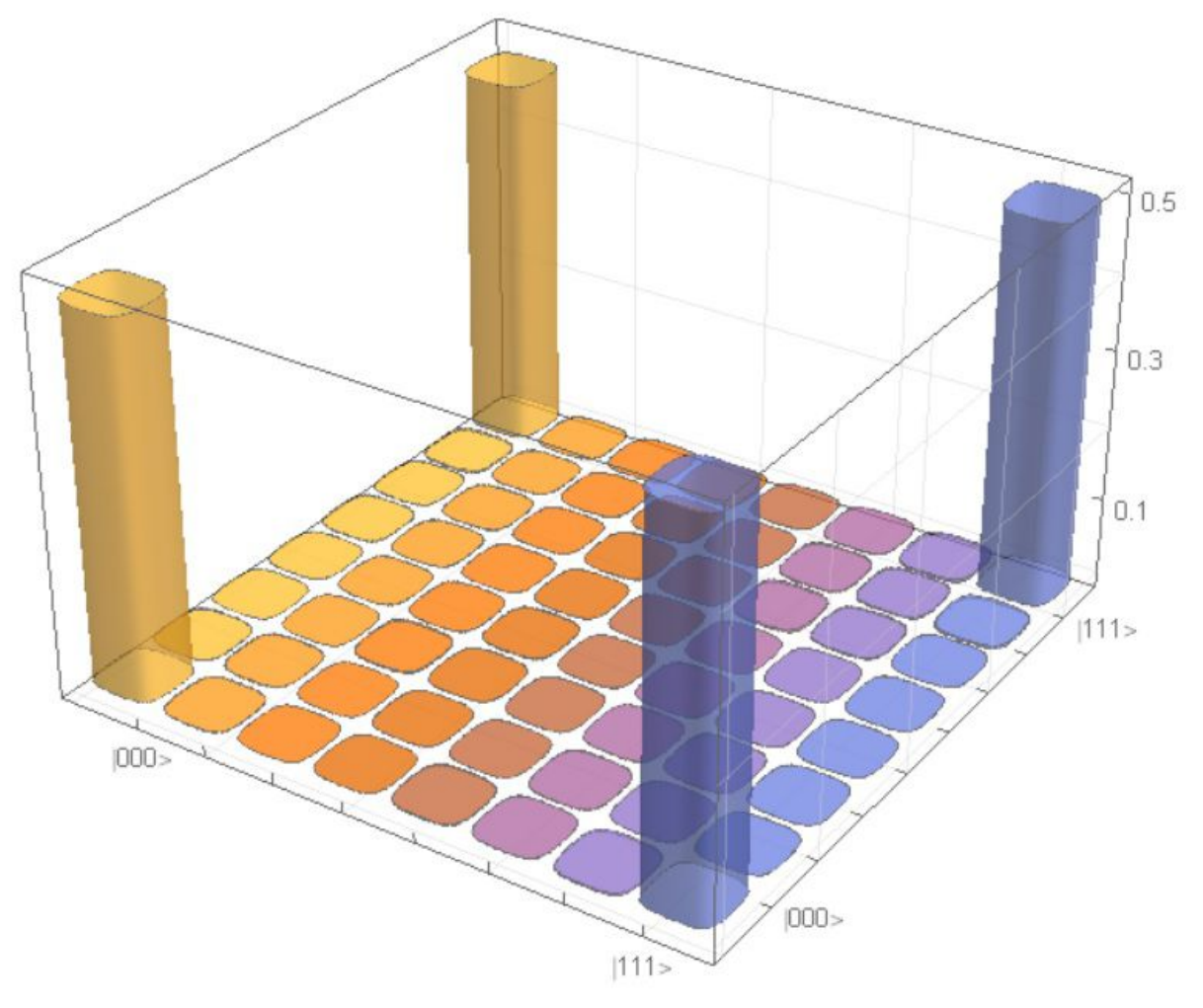}}\end{tabular}&
			\begin{tabular}{l}\raisebox{- \height}{\includegraphics[width=0.65\columnwidth]{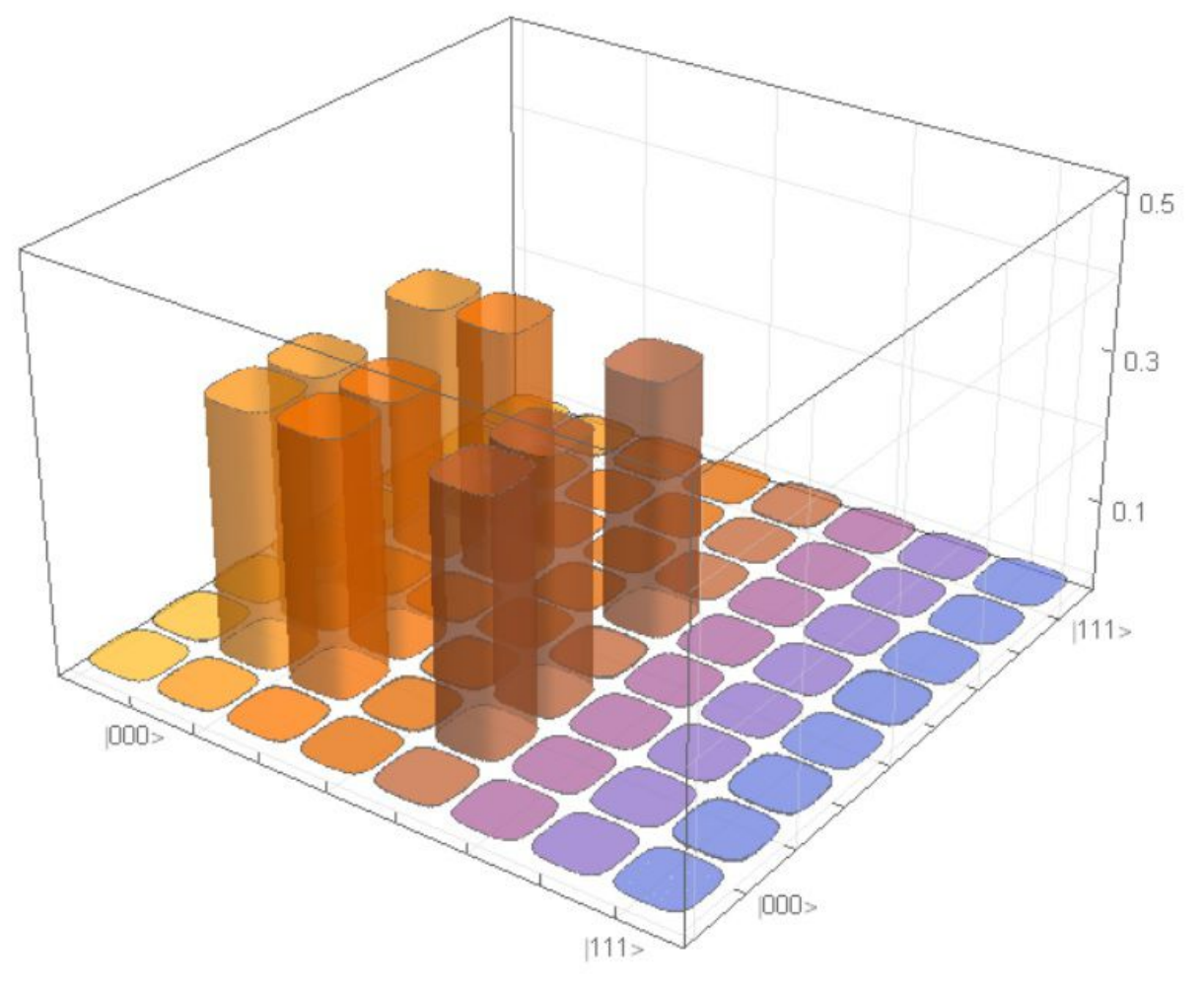}}\end{tabular}\\
\hline
			\centered{$100$}&\begin{tabular}{l}\raisebox{- \height}{\includegraphics[width=0.65\columnwidth]{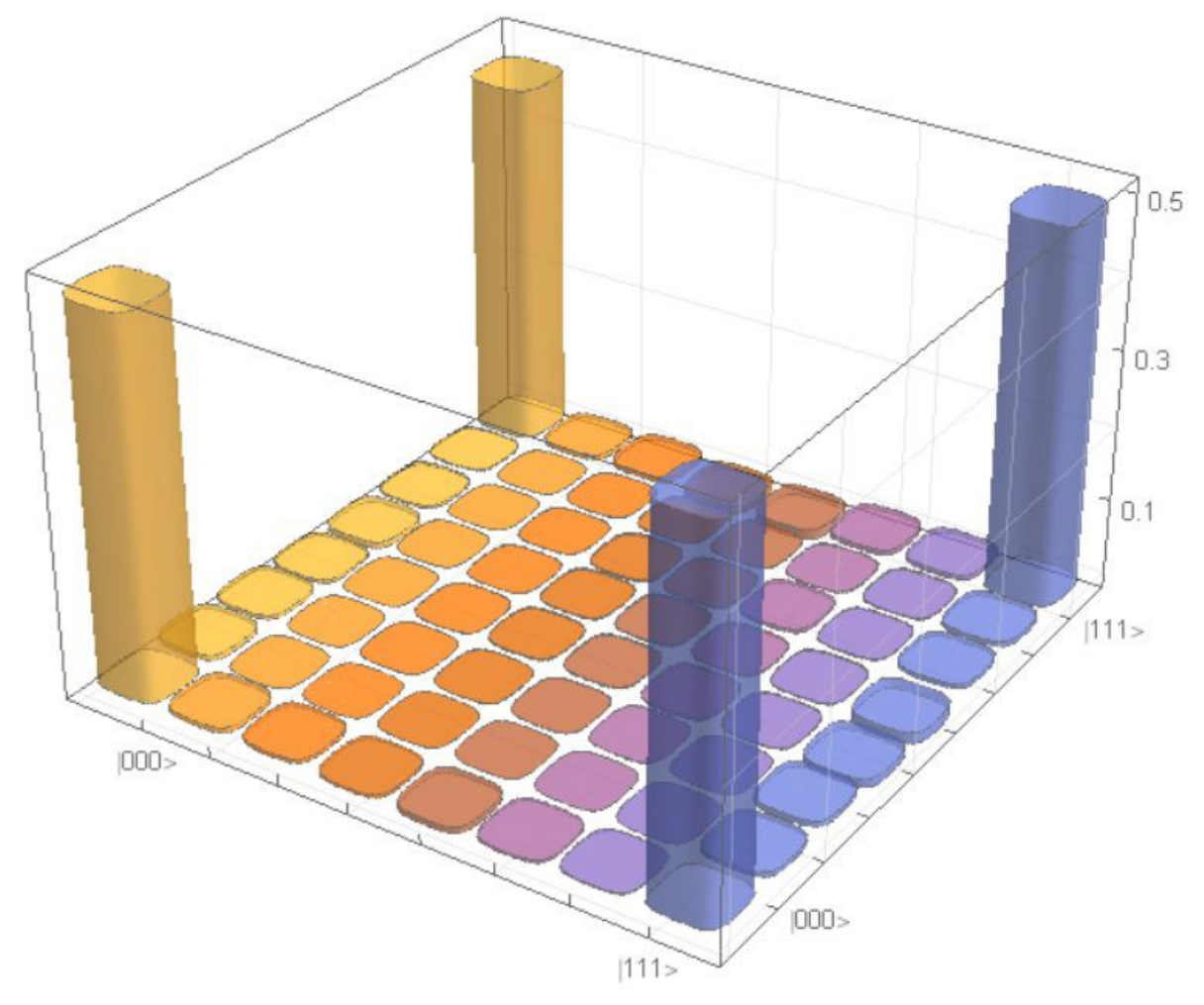}}\end{tabular}&
			\begin{tabular}{l}\raisebox{- \height}{\includegraphics[width=0.65\columnwidth]{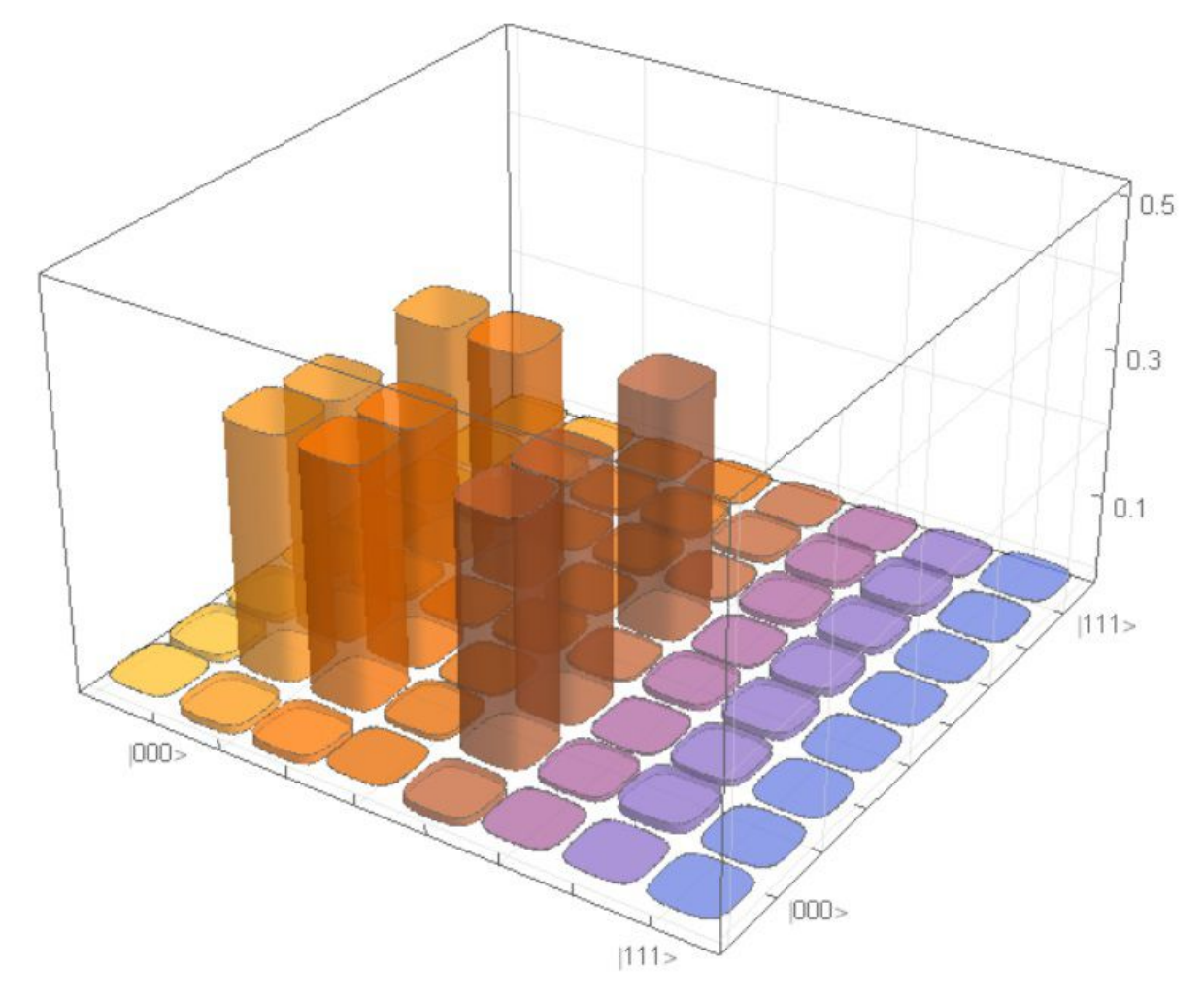}}\end{tabular}\\
\hline
			\centered{$10$}&\begin{tabular}{l}\raisebox{- \height}{\includegraphics[width=0.65\columnwidth]{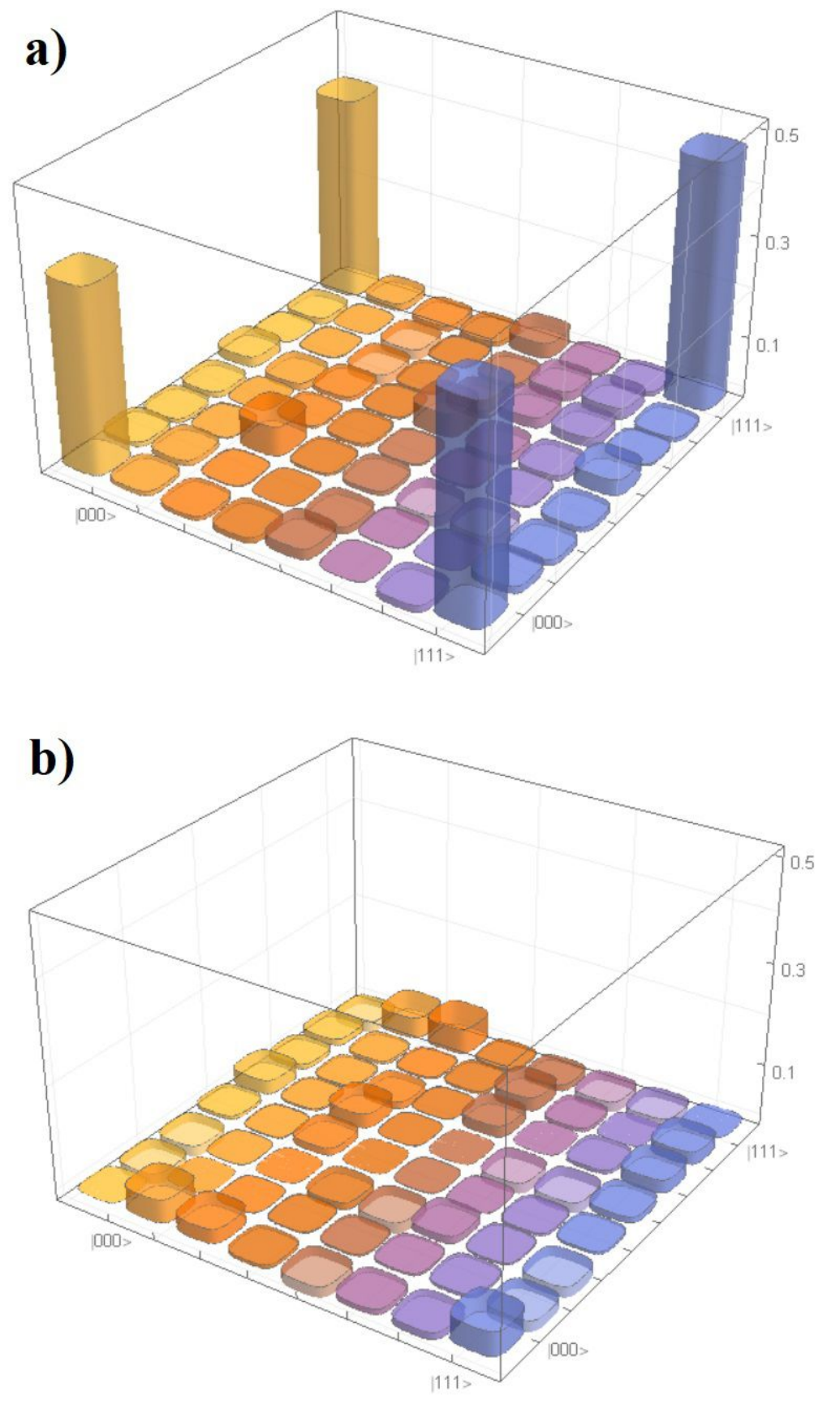}}\end{tabular}&
			\begin{tabular}{l}\raisebox{- \height}{\includegraphics[width=0.65\columnwidth]{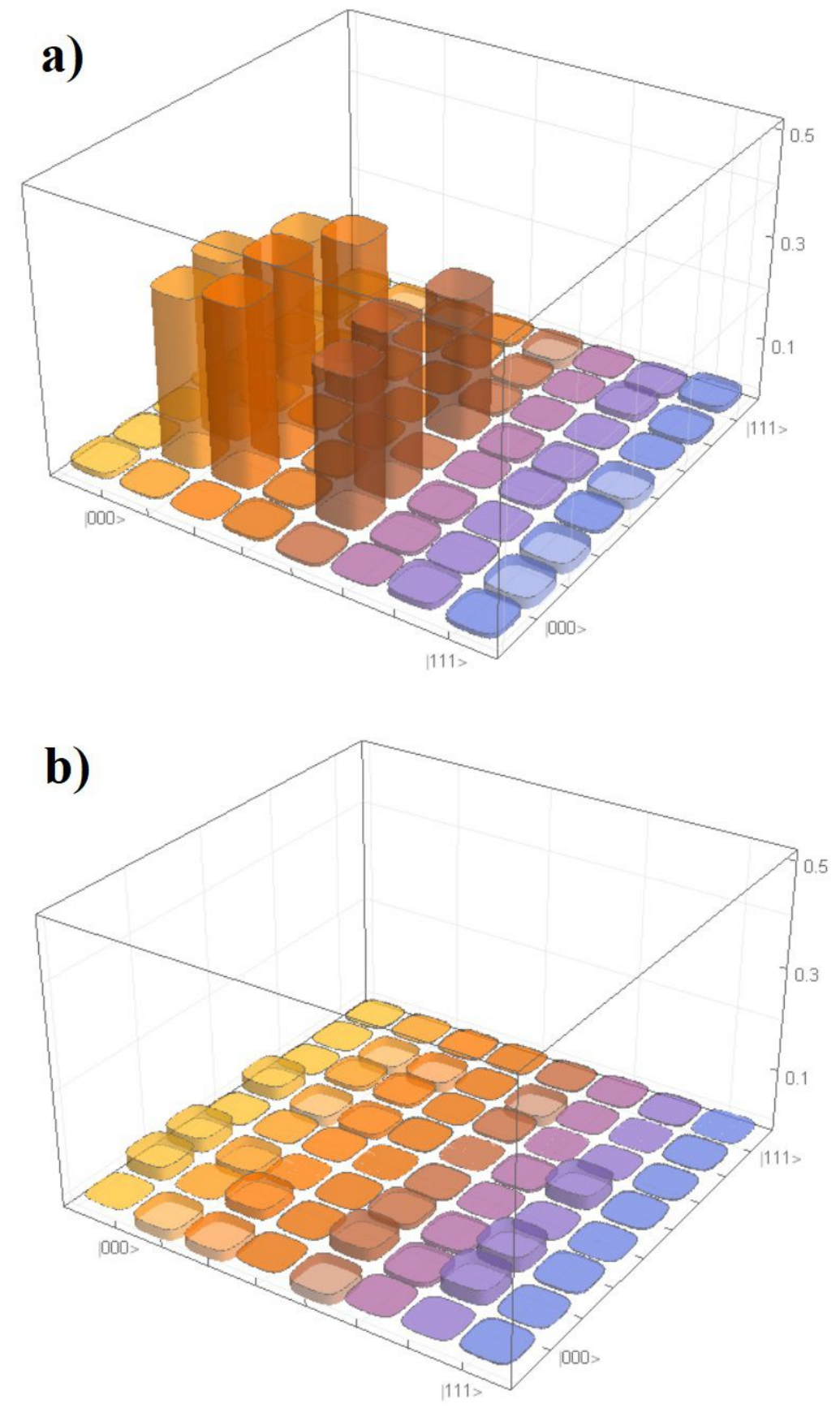}}\end{tabular}\\
\hline
		\end{tabular}
		\caption{Density matrices obtained from the framework for two input states: $\ket{GHZ}$ and $\ket{W}$. For $\mathcal{N}=10$, the panels show real (a) and imaginary (b) parts of the reconstructed state.}
		\label{reconstruction}
	\end{figure*}

As already indicated, for $\eta=1$, the quantum state \eqref{w4} corresponds to the GHZ state. Since $\ket{GHZ}$ plays a crucial role in quantum information theory, it appears justifiable to consider this special case separately. In \figref{reconstruction}, one finds, in the left-hand side column, the reconstructed state for three selected ensemble sizes. In the case of $\mathcal{N} = 1\,000$ and $\mathcal{N} = 100$, we present only the real part of the estimated density matrix because the imaginary part is negligible. For $\mathcal{N} = 10$, we visualize both real and imaginary parts to show inaccuracies of state estimation in this scenario. Similarly, in the right-hand side column of \figref{reconstruction}, one has the results of applying the QST framework to the W state.

First, for $\mathcal{N} = 1\,000$, we notice that both $\ket{GHZ}$ and $\ket{W}$ can be accurately reconstructed by the framework. The fidelity between $\ket{GHZ}$ and its estimate is $0.9934$. For $\ket{W}$ the overlap equals $0.9997$. This proves that the framework is efficient not only for three-qubit generalized Werner states but also for the W state, which features a different kind of entanglement. The result is in line with the above analysis and confirms that the ensemble size $\mathcal{N} = 1\,000$ guarantees perfect accuracy of state reconstruction.

If $\mathcal{N} = 100$, one can spot minor flaws in the results of state estimation. It can be expressed numerically by the quantum fidelity, which equals $0.9540$ for the GHZ state estimation and $0.9682$ in the case of the W state. One can notice that the accuracy of the framework is slightly better for the W state.

Finally, if $\mathcal{N} = 10$, we observe a considerable amount of errors in the reconstructed states. In particular, the state tomography of $\ket{GHZ}$ yields the fidelity equal $0.8216$. Errors are noticeable not only in the real part of the reconstructed density matrix but also in the imaginary part, which should be flat. It appears that the Poisson noise is less detrimental for W state estimation, which results in the fidelity: $0.9424$. To conclude, we can state that the framework seems more robust against the Poisson noise if it is applied to the W state reconstruction rather than the GHZ state.

\subsection{Coherence of the GHZ state}

\begin{figure}[h]
	\centering
         \centered{\includegraphics[width=0.75\columnwidth]{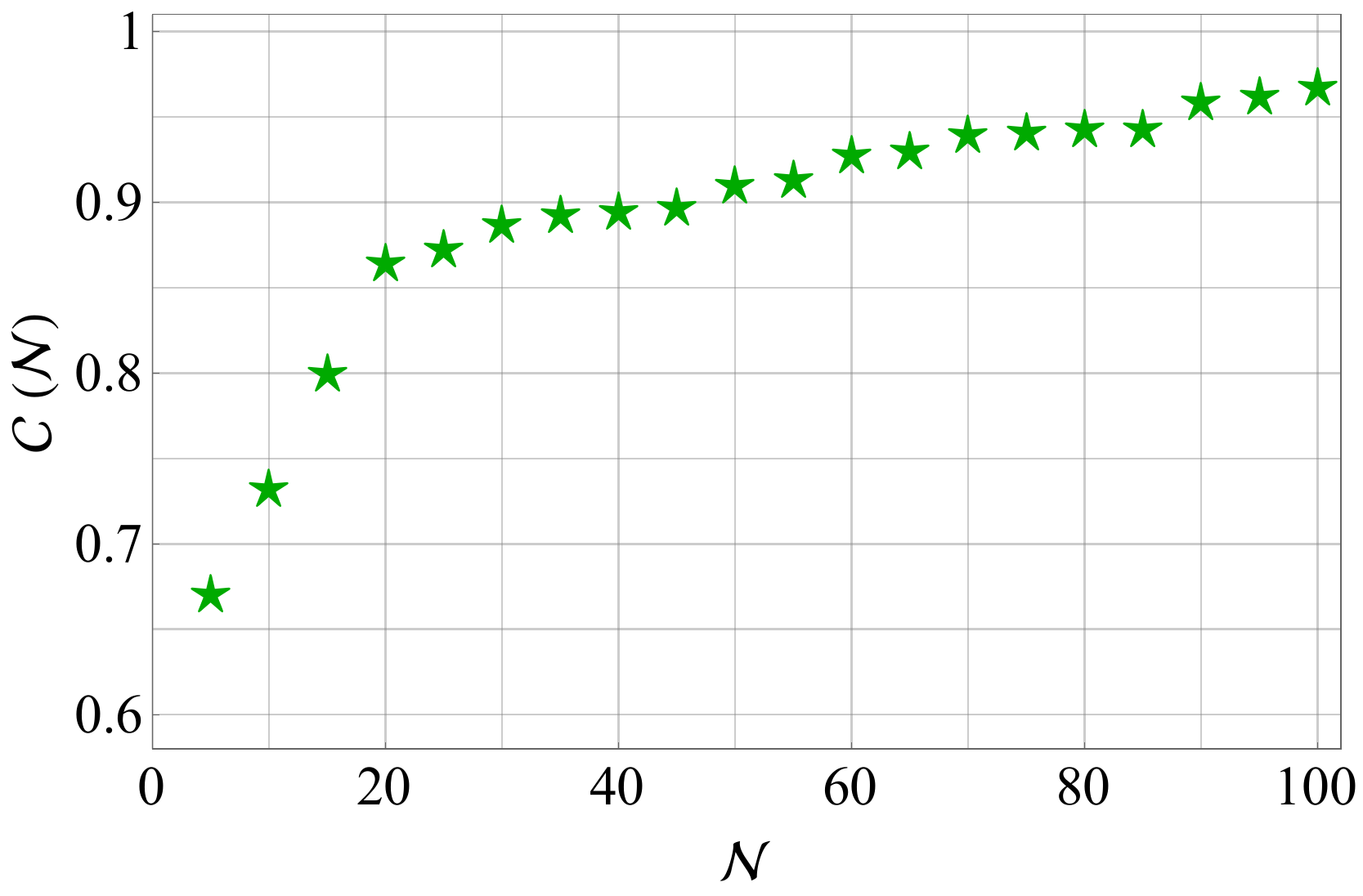}}
	\caption{Dots present the coherence, $\mathcal{C} (\mathcal{N})$, of reconstructed states obtained from the tomographic framework. In each case, the GHZ state was the input state.}
	\label{coherence}
\end{figure}

In practical applications, the quality of GHZ states is quantified in time by computing its coherence \cite{Monz2011,Ozaeta2019}. This approach allows one to track how the state degenerates on account of interactions between the system and its environment. The coherence, denoted by $\mathcal{C}$, is defined as the sum of the amplitudes of the far-off-diagonal elements of the density matrix representing the state. Thus, for the actual GHZ state, the coherence reaches its maximum value, which is one. In our framework, the measured state is affected by the Poisson noise, which impacts its coherence. Assuming that the GHZ state is the input state for the framework, we can consider the coherence of reconstructed states as a function of the ensemble size, which is denoted by $\mathcal{C} (\mathcal{N})$, and presented in \figref{coherence}.

First, we notice that $\mathcal{C} (\mathcal{N})$ is a monotone function that increases with the ensemble size. This observation was anticipated since the ensemble size is positively correlated with the accuracy of state tomography, which is expected to cause a boost of coherence. However, the exact shape of the function appears more intriguing. The coherence is a concave function of the ensemble size. In particular, one can observe a rapid growth of $\mathcal{C} (\mathcal{N})$ for the first few arguments. Such an analysis allows one to determine before an experiment the ensemble size necessary to obtain a specific value of coherence.

\section{Discussion and summary}\label{discussion}

In the article, we introduced a mathematical model for three-qubit quantum state tomography, which is based on the SIC-POVM and involves the Poisson noise as a source of experimental uncertainty. By numerical simulations, we tested the framework on three-qubit generalized Werner states. In particular, the results have revealed that if we utilize $1000$ photon pairs per measurement, the framework is robust against the Poisson noise. In this case, the quantum fidelity is, approximately, a constant function with the value close to one, which ensures high-accuracy state tomography along the entire domain of three-qubit generalized Werner states.

Furthermore, we discovered that if we exploit only $10$ copies of the system per measurement, the framework is vulnerable to the Poisson noise, which results in poor fidelity. However, the accuracy of state estimation improves as we reduce the entropy of the input states.

As a special case, we investigated the performance of the framework on the GHZ state, which corresponds to the pure three-qubit generalized Werner state. The accuracy of the tomographic technique was compared with the estimation of the W state, which is considered a distinct kind of three-qubit entanglement. It was demonstrated that the framework is more accurate at reconstructing the W state, which was particularly evident for the small ensemble size. Thus, we can conclude that the two representatives of three-qubit entanglement, i.e., $\ket{GHZ}$ and $\ket{W}$, differ in the quality of state tomography under the Poisson noise.

Furthermore, particular attention was paid to quantum tomography of the GHZ state versus the ensemble size. The density matrix obtained from the QST framework can only resemble the input state due to the noise that distorts the measurements. Thus, we investigated how well the coherence is preserved under the noisy conditions. Through numerical simulations, the interdependence between the coherence and the ensemble size was determined.

The findings of the article appear to have practical implications that can facilitate experimental quantum state tomography. More specifically, one can numerically determine the minimal ensemble size required to obtain a requested level of quality. If an experimenter knows the purity of the state produced by the source, the ensemble size can be learned precisely. Since the time of photon detection is correlated with the ensemble size, this allows one to optimize the experiment.

In the future, the framework will be tested on other classes of multipartite entangled states. Additionally, alternative kinds of experimental noise will be incorporated into the model to study the robustness of the method in different scenarios.

\section*{Acknowledgments}

The research was supported by the National Science Centre in Poland, grant No. 2020/39/I/ST2/02922.

\end{document}